\documentclass[10pt]{article}
\usepackage[letterpaper]{geometry}
\usepackage{hicss}
\usepackage{times}
\usepackage[none]{hyphenat}
\usepackage{url}
\usepackage{latexsym}
\usepackage{minted}
\usepackage{indentfirst}
\usepackage{graphicx}
\usepackage{comment}
\graphicspath{{images/}}
\usepackage[style=apa,]{biblatex}
\usepackage{tabu}                      
\usepackage{booktabs}                  
\usepackage{lipsum}                    
\usepackage{mwe}                       
\usepackage{amssymb}
\usepackage[normalem]{ulem}
\usepackage{mathptmx}
\usepackage[subtle]{savetrees}
\usepackage{subcaption}
\usepackage{fancyhdr}

\title{Climate Data for Power Systems Applications: Lessons in Reusing Wildfire Smoke Data for Solar PV Studies}

 \author{Arleth Salinas \\
  University of Utah \\
  {\underline{ arleth.salinas@utah.edu}} \\ \\
  Saud Amjad \\
  University of Calgary \\
  {\underline{ saud.amjad@ucalgary.ca} } \\ \\
  Nicolas Duboc \\
  University of British Columbia \\
  {\underline{ nduboc@eoas.ubc.ca} } \\ \And
  Irtaza Sohail \\
  University of Calgary \\
  {\underline{ irtaza.sohail@ucalgary.ca} } \\ \\
  Aashish Panta\\
  University of Utah \\
  {\underline{ aashish.panta@utah.edu} } \\ \\
  Mostafa Farrokhabadi \\
  University of Calgary \\
  {\underline{ dgri@ucalgary.ca} } \\ \And
  Valerio Pascucci \\
  University of Utah \\
  {\underline{ pascucci@sci.utah.edu} } \\ \\
  Roland Schigas\\
  University of British Columbia \\
  {\underline{ roland.schigas@ubc.ca} } \\\\
  Roland Stull\\
  University of British Columbia \\
  {\underline{ roland.stull@ubc.ca} } \\ \And
  Pantelis Stefanakis \\
  University of Calgary \\
  {\underline{ pantelis.stefanak1@ucalgary.ca} } \\ \\
  Timothy Chun-Yiu Chui\\
  University of British Columbia \\
  {\underline{ tchui@eoas.ubc.ca} } \\}

\date{}

\begin{document}

\thispagestyle{fancy}
\fancyhf{} 
\renewcommand{\headrulewidth}{0pt} 
\renewcommand{\footrulewidth}{0pt} 
\lfoot[]{\it{This paper has been accepted for the upcoming 59th Hawaii International Conference on System Sciences (HICSS-59).}}

\maketitle
\begin{abstract}
Data reuse is using data for a purpose distinct from its original intent. As data sharing becomes more prevalent in science, enabling effective data reuse is increasingly important. In this paper, we present a power systems case study of data repurposing for enabling data reuse. We define data repurposing as the process of transforming data to fit a new research purpose. In our case study, we repurpose a geospatial wildfire smoke forecast dataset into a historical dataset. We analyze its efficacy toward analyzing wildfire smoke impact on solar photovoltaic energy production. We also provide documentation and interactive demos for using the repurposed dataset. We identify key enablers of data reuse including metadata standardization, contextual documentation, and communication between data creators and reusers. We also identify obstacles to data reuse such as risk of misinterpretation and barriers to efficient data access. Through an iterative approach to data repurposing, we demonstrate how leveraging and expanding knowledge transfer infrastructures like online documentation, interactive visualizations, and data streaming directly address these obstacles. The findings facilitate big data use from other domains for power systems applications and grid resiliency.
\end{abstract}

\subsubsection*{Keywords:}

Climate Data, Grid Resiliency, Wildfire, Solar PV, Data Reuse

\section{Introduction}

Big Data is becoming increasingly important for power systems research and development. In the broader context of science and engineering, governments have begun increasing data sharing requirements \parencite{OECDDeclaration, NationalAcademiesOpenScience}. Researchers seek credit for their shared data \parencite{SilvelloDataCitation}. Shared data is increasingly discovered by and processed by machines for artificial-intelligence powered research and other data-intensive research processes \parencite{10505129}. 

These forces are creating a deluge of scientific data \parencite{BorgmannConundrum}. Researchers who create and publish data do not always have the bandwidth to ensure their data are well-prepared for integrative reuse \parencite{WilkinsonFAIR}. Integrative reuse includes running new analyses, incorporating shared data into other data sources, and building upon shared datasets \parencite{BorgmannConundrum}. This is of particular importance in power systems research given the tight coupling with other physical domains, including climate, transportation, thermal systems, etc. Therefore, data repurposing is needed to enable data reuse in power systems. We define data repurposing as the transformation of data to fit a new research purpose that is distinct from its original purpose. Data repurposing can be difficult due to barriers to data access, lack of documentation, and risk of misinterpretation \parencite{YoonRedFlags}.

In this paper, we provide a case study in data repurposing for enabling the reuse of the geospatial wildfire smoke forecast dataset (herein referred to as the FireSmoke dataset) from the BlueSky Canada (BSC) project. The FireSmoke dataset is often repurposed in atmospheric science applications. However, repurposing this dataset as a non-domain expert presents obstacles to its reusability. In order to enhance the reusability of this scientific dataset we ask what difficulties does one encounter in the repurposing of scientific data, and what benefits are yielded? Our contributions are:
\begin{itemize}
    \item Highlighting challenges and opportunities in enabling data repurposing of scientific data based on our experience repurposing the FireSmoke dataset.
    \item A proof-of-concept of scientists reusing the repurposed dataset in a novel analysis on the effect of wildfire smoke on solar photovoltaic (PV) energy production efficacy.
    \item Public access to the repurposed FireSmoke dataset published on the cloud, along with documentation and interactive demos.
\end{itemize}

\section{Wildfire Smoke and Solar PV Data}
Wildfire events are on the rise due to a changing climate \parencite{FireRegime}. In 2023 the carbon emissions from Canadian wildfires were only next to the total annual carbon emissions of the US, China and India \parencite{Nature}. Simultaneously, solar PV installations are expanding as part of global efforts to mitigate climate change. By 2030, solar PV production is projected to triple, surpassing wind and hydroelectric power to become the dominant renewable electricity source \parencite{IEA}.

The concurrent rise in wildfire incidents and large-scale solar PV deployment has raised concerns regarding the impact of wildfire smoke on solar energy generation. To assess the effect of wildfire smoke on solar PV performance, an appropriate metric for quantifying smoke must first be identified. Biomass combustion, including forest fires, primarily produces fine particulate matter smaller than 2.5 $\mu$m (PM2.5) \parencite{Biomass}. Thus, PM2.5 is commonly used in research as a proxy for wildfire smoke.

Previous research relies on distant PM2.5 sensors. For example \textcite{GILLETLY} analyzed the impact of wildfire smoke on solar PV production across 53 locations in the western United States, using PM2.5 measurements from stations up to 48 km away. Some solar PV sites were excluded due to the lack of sufficiently close measurement stations. The authors acknowledged that the spread of PM2.5 from wildfires was affected by distance and orientation (wind direction, topography, etc.) of measurement locations.

\textcite{Korea} studied the effect of particulate matter, including PM2.5 and PM10, on solar photovoltaic power generation. The nearest PM2.5 measurement stations were 8.8 km and 4.5 km from the solar photovoltaic installations. The authors acknowledged that relying on distant sensors introduced uncertainties, noting that more localized PM2.5 values would have strengthened their analysis.

Even when monitoring stations are relatively close inaccuracies may arise if factors such as wind speed and direction are not considered. This underscores the challenge of using ground-based PM2.5 data for assessing wildfire smoke impacts on solar PV output.

To address these limitations, some studies have employed alternative metrics, such as Aerosol Optical Depth (AOD), to measure wildfire smoke. For example, \textcite{Donaldson} examined the impact of wildfire smoke in California using AOD data from a geostationary satellite with a spatial resolution of 2 km. However, satellite-based AOD measurements are limited to clear and snow-free conditions and can be influenced by non-wildfire aerosols. Due to these data quality concerns, some researchers, such as \textcite{GILLETLY}, have opted against using AOD data in their analyses.

Other researchers have circumvented direct smoke quantification and instead analyzed the impact of specific wildfire events on solar PV performance. For instance, \textcite{Australia} investigated the effects of a wildfire in Australia without quantifying the smoke concentration. While such studies can provide some valuable insights, they cannot be directly compared with other studies due to the absence of quantified smoke data.

\section{FireSmoke Dataset}
Under typical data repurposing, researchers may not be in direct collaboration with the original data creators. In this section, the data creators from the University of British Columbia (UBC) describe the FireSmoke dataset for completion. In Section 4, we discuss repurposing the data with minimal input from UBC, to illustrate the data repurposing process for this specific application.

\subsection{Data Creation}
The BlueSky Canada (BSC) dataset analyzed in this study is published through the FireSmoke portal (firesmoke.ca), which provides high-resolution forecasts of wildfire smoke dispersion across North America.

To create smoke dispersion forecasts, satellite hotspot detects must be obtained to determine fire locations and intensity, and a North America-wide high-resolution weather model must first be run to generate the meteorological fields to transport the pollutants. The workflow for producing wildfire smoke forecasts using the BlueSky Canada Pipeline (BSP) modeling framework involves the following:

\begin{itemize}
\item Download the North American Mesoscale (NAM) model from the National Centers for Environmental Prediction (NCEP) to provide the initial and boundary conditions to run the high-resolution weather model.
\item Run the Weather Research and Forecasting (WRF) model version 4.2.1 \parencite{Skamarock_Klemp_Dudhia_Gill_Liu_Berner_Wang_Powers_Duda_Barker_et} using the NAM data, to generate meteorological fields as input into the dispersion model. WRF is a non-hydrostatic weather model that supports domain nesting to increase horizontal spatial resolution, and this feature is used to produce weather forecasts with a horizon of 84 hours using a 36-12 km nested domain configuration.
\item Download real-time wildfire location and fuel consumption estimates from the Canadian Wildland Fire Information System (CWFIS) \parencite{Canadian_wildfire}, to determine fire hotspots and to estimate emissions.
\item Cluster and associate fire hotspots into a reconciliation stream, used as input to BSP persistence, growth, and plume rise models. Obtain ``carryover" smoke from previous dispersion runs to ensure continuity of smoke concentrations across simulations and to reduce the likelihood of low biases.
\item Combine the meteorological fields from the WRF output on the inner nest with 12-km grid spacing; the fire location and emissions data from the BSP models; and the carryover smoke from prior simulations, to run the HYSPLIT dispersion model \parencite{Stein_Draxler_Rolph_Stunder_Cohen_Ngan_2015}. HYSPLIT is an atmospheric dispersion model that supports a variety of simulation configurations, including the "puff" mode chosen to produce efficient real-time simulations for BlueSky Canada. The output of HYSPLIT is in NetCDF format, and contains hourly PM2.5 concentrations.
\end{itemize}

This dataset is projected onto a structured latitude-longitude coordinate system with a grid spacing of 0.1° $\times$ 0.1° ($381 \times 1081$ grid points), allowing for fine-grained assessments of smoke-related air pollution over North America. The geographic domain extends from -160.0° to -66.0° longitude and 32.0° to 51.0° latitude, covering a substantial portion of North America. The primary variable of PM2.5 (in µg/m³) represents the surface concentrations of pollutants across the domain.

These smoke forecasts are provided 4-times daily during the Canadian wildfire season (March - October), resulting in overlapping validity periods with updates to the forecast every 6 hours. Because weather and smoke forecasting are initial-value problems, the most updated forecast is generally considered to be the most skillful, though verification against ground observations of pollutant concentrations is ongoing.

\subsection{The NetCDF Data Format}
The FireSmoke portal (firesmoke.ca) provides PM2.5 concentrations in Network Common Data Form (NetCDF) files \parencite{NetCDF}. NetCDF was developed by the University Center for Atmospheric Research (UCAR), and is a gridded data format commonly used in the atmospheric, oceanic, and climate sciences. It is also widely used in wildfire smoke research to store pollutant concentration data, which can be used for air quality analyses. Such analyses are often conducted in the context of public health and disaster management, particularly during the Canadian wildfire season. Because these datasets are often forecasts with multiple initialization times that have overlapping validity periods over the same spatial area, one must account for redundant or conflicting values to ensure data consistency and prevent data duplication.

The high-dimensional nature of NetCDF files can also pose computational challenges. Geospatial data are often multidimensional across time, latitude, longitude, and vertical levels for three-dimensional variables, and can therefore result in high demands on storage, memory, and I/O performance. Given data volumes on the order of hundreds of GB to TB, across potentially thousands to hundreds of thousands of files, scalable solutions such as progressive streaming and parallel computational approaches are often necessary to facilitate real-time and large-scale analyses. Utilities such as the Xarray package in Python can support the processing of these files at scale and are leveraged in this paper \parencite{Hoyer_Hamman_2017}.

\section{Data Repurposing Process}
In this section, we describe the repurposing of the FireSmoke dataset into a historical dataset as led by computer science researchers at the University of Utah. We highlight the challenges a non-atmospheric scientist may face when repurposing the dataset, as well as the strengths of the knowledge infrastructures provided by UBC for enabling this process.

\subsection{Data Loading}
During the fire season, forecasts are done 4-times a day (at 0, 6, 12, and 18 Universal Standard Time, hereby referred to as UTC) with 12 km nests focusing on the United States and Canada. Each forecast has a horizon of 84 hours (3.5 days). The forecasts are published as NetCDF files and are given a forecast ID according to the time of their publication: BSC00CA12-01, BSC06CA12-01, BSC12CA12-01, BSC18CA12-01. We chose to download all available NetCDF files from the FireSmoke portal.


\subsubsection{Downloading NetCDF Files}

\begin{figure}[ht]
    \centering    \includegraphics[width=0.75\linewidth]{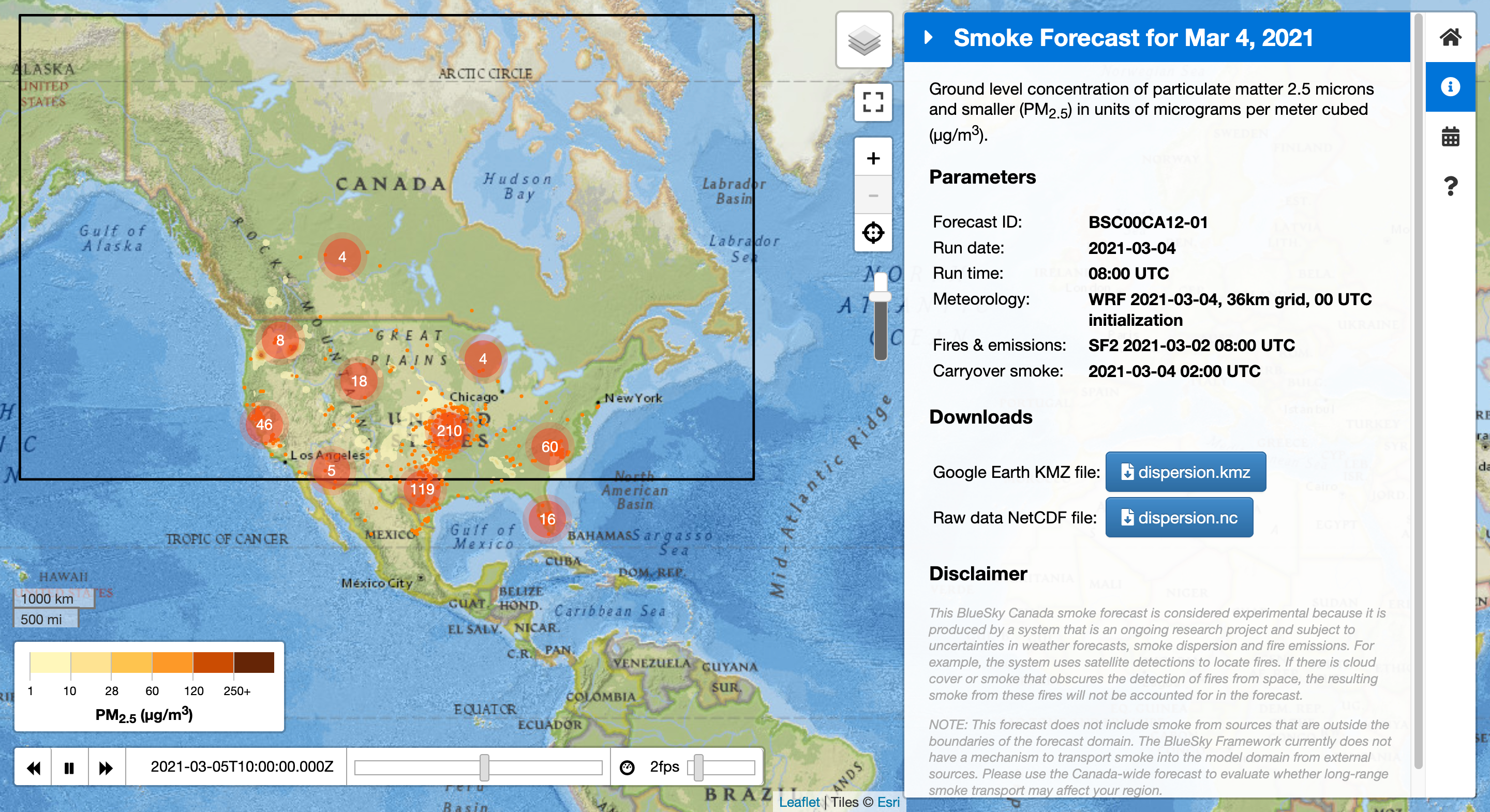}
    \caption{The BlueSky Canada forecast for March 4, 2021 on the FireSmoke portal.}
    \label{fig:firesmoke_portal}
\end{figure}

The FireSmoke portal provides a geospatial visualization of the latest forecast and download links for the NetCDF file or KMZ file as shown in Figure \ref{fig:firesmoke_portal}. 
 
The NetCDF file for a specific forecast published on a specific date can be accessed by navigating to the corresponding URL:
\texttt{\path{https://firesmoke.ca/forecasts/{Forecast ID}/{YYYYMMDD}{InitTime}/dispersion.{FileExtension}}}

"Forecast ID" and "InitTime" are populated using the desired forecast ID and publication time in UTC, "YYYYMMDD" is the publication date of ones choice, and "FileExtension" is either ".nc" or ".kmz".

UBC provided these instructions, which are not published on the FireSmoke portal. An outside user would need to extrapolate these by inspecting various URLs on the FireSmoke portal for each forecast.

\subsubsection{Determining Available Dates}

The dates for which forecast data are available for download are not published. Knowing the project began in 2021, we inferred that the earliest available date would be in 2021.

We navigated to various URLs in a browser using the instructions mentioned previously, and found the earliest published forecast to be in March of 2021. We used wget to download the file at at the corresponding URL for each forecast ID for every day from March 3, 2021 to June 27, 2024 (the last time we ran this download script). There are about 1000 NetCDF files for each forecast ID.

Collaborating directly with UBC gave us information that is otherwise unpublished. UBC informed us that they store FireSmoke data created earlier than 2021 in tape storage to be more cost efficient. UBC also informed us that recent forecast data would be stored to tape, causing us to soon lose access to recent forecast data. This prompted us to download the most recent data published after June 2024.


\subsection{Initial Exploration}
We familiarized ourselves with the FireSmoke dataset by creating a video of rudimentary visualizations of all the hours from March 2021 to June 2024. At this point, we only used the BSC00CA12-01 forecast. This process revealed various issues.

\subsubsection{Metadata Understanding}

We used Xarray view the NetCDF files. As non-experts in the development of the dataset and atmospheric science, interpreting the metadata within the files was unclear. We searched for resources explaining how to use the metadata to compute the latitude and longitude coordinates to plot the PM2.5 concentration data.

\subsubsection{Failed Downloads}

We appeared to successfully download all forecasts over all dates specified when using wget. However, many files failed to open. We realized wget downloaded the contents of an HTML webpage to a dispersion.nc file when a true dispersion.nc file was unavailable at the specified URL. We had to adjust our downloading script and verify that the URL indeed pointed to a NetCDF file by inspecting the headers of the URL first.

After fixing this, we still noticed many files failed to open. We identified the files that failed to open with Xarray and cross-checked them against empty plots in our video, confirming the presence of missing data. We discovered that files were either corrupt or not downloaded at all. Thus, we re-downloaded the files that Xarray failed to open from the FireSmoke portal. We cross-checked all the timestamps in the newly downloaded files with the timestamps that were visualized as empty in our curated dataset. We concluded that a subset of files truly failed to download, and we confirmed that some files were truly unavailable. The FireSmoke portal clarifies that sometimes forecasts fail to run, resulting in no file available.

\subsubsection{Grid Size Miscommunication}

Our initial visualizations showed nonsensical results, such as wildfire smoke originating from the Pacific Ocean.

We discovered that the data from these nonsensical visualizations used a grid of size $381 \times 1041$ columns upon inspecting the metadata in the NetCDF file. UBC reported that the dataset used a consistent grid of $381 \times 1081$ across all files. Therefore, we used the same size grid and plotting routine for all time steps across all files. We never checked whether the metadata across all NetCDF files was the same.

We compared all metadata across the files and discovered that the earliest month of available data used a $381 \times 1041$ grid size. Therefore, in order to create a physically reasonable visualization, we used interpolation to fit the data onto a $381 \times 1081$ grid. However, when we shared this approach with UBC, we were informed that it is more appropriate to use the data in the original $381 \times 1041$ domain. This is because the forecast created in the $381 \times 1041$ domain is not interchangeable with the forecast that would have been created using the $381 \times 1081$ domain.

\subsection{Data Curation}
After identifying various pitfalls through our initial exploration, we created the historical dataset. Our approach had three stages: a) check which NetCDF files were successfully downloaded from the FireSmoke portal; b) select a subset of data from the collection of forecast data to create a chronological, hour-by-hour dataset; c) save the dataset to an IDX file using the OpenVisus framework and save the metadata to a NetCDF file.

We use the IDX file format to store the curated dataset and a NetCDF file to hold the original metadata of the FireSmoke data, derivative metadata from curating the FireSmoke data, and the public URL to the IDX file. Figure \ref{fig:quarto1} shows the final NetCDF file open in Xarray in our Quarto documentation which is available in our supplemental materials.

The IDX file format enables data loading at different resolutions and at specified time steps to avoid loading all of the data into memory. This enables interactive data exploration and usage despite potential memory limits from a user machine \parencite{10767643, Kumar2014}.


\subsubsection{Validating Downloaded NetCDF Files}

All downloaded files are organized into directories by their forecast ID. We append the creation date of each file to the file name. For example, a file generated on March 4, 2021 with forecast ID BSC00CA12-01 is named \texttt{dispersion\_20210304.nc} and stored in directory \texttt{BSC00CA12-01}.

Then we check for the following for all downloaded files:
\begin{enumerate}
    \item Ensure sure all downloaded files open successfully with Xarray.
    \item Ensure consistency of metadata across all files.
\end{enumerate}
\subsubsection{Sequencing of Hourly Data}

\begin{figure}[ht]
    \centering
    \includegraphics[width=\linewidth]{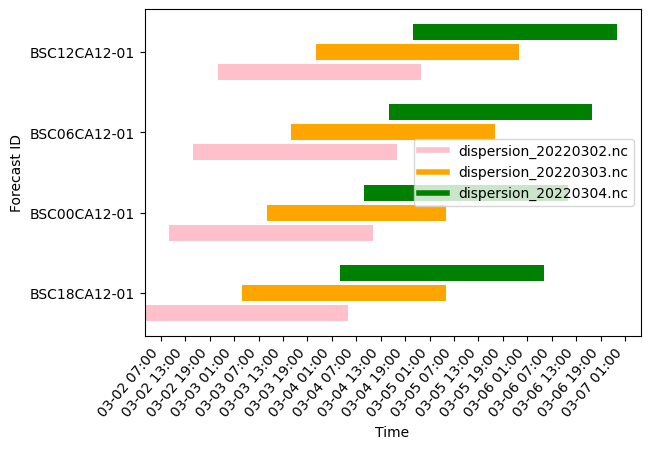}
    \caption{Time horizons for all forecasts generated on March 2-4 2022, grouped by forecast ID.}
    \label{fig:stagger}
\end{figure}

Next, our script selects the last updated forecast for a given time. There may be a variable number of files for which a forecast is given, for any given time step. For example, the latest forecast for March 4, 2022 01:00 UTC would be in the \texttt{dispersion.nc} file published on March 4, 2022 with forecast ID BSC00CA12-01. However, older forecasts are provided as well, both in BSC00CA12-01 and in the three other forecast ID directories. Figure \ref{fig:stagger} demonstrates the staggered nature of the forecast data. Because files are arbitrarily missing in the FireSmoke dataset, we had to implement the logic for selecting the latest forecast accordingly.

\begin{figure}[ht]
    \centering
    \begin{subfigure}[b]{0.48\linewidth}
        \centering
        \includegraphics[width=0.75\linewidth]{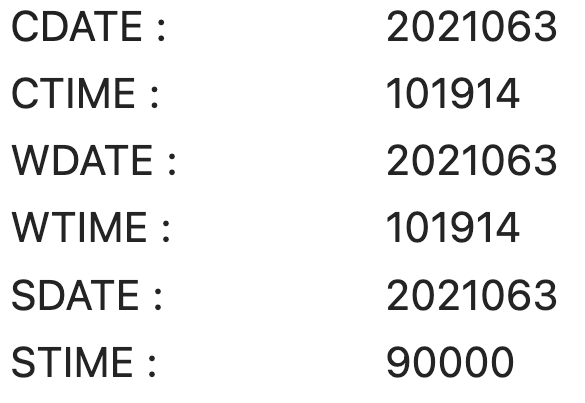}
        \caption{Metadata from the NetCDF file.}
        \label{fig:metadata-xarray}
    \end{subfigure}
    \hfill
    \begin{subfigure}[b]{0.48\linewidth}
        \centering
        \includegraphics[width=0.75\linewidth]{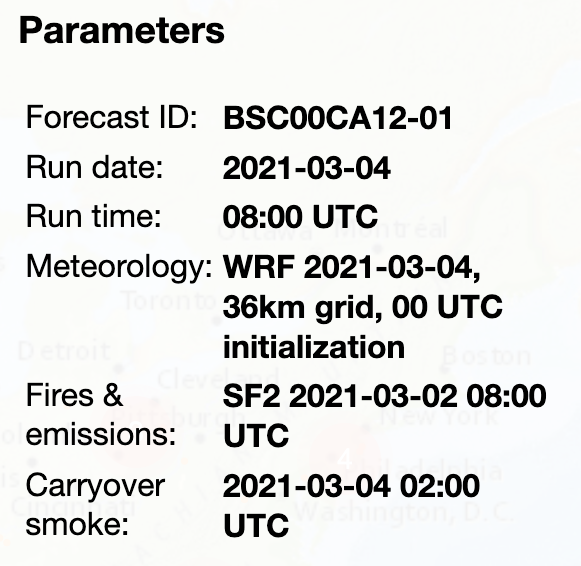}
        \caption{Metadata from the FireSmoke portal.}
        \label{fig:metadata-firesmokeportal}
    \end{subfigure}
    \caption{Metadata for March 4, 2021 forecast with forecast ID BSC00CA12-01.}
    \label{fig:metaportal}
\end{figure}

First, we identified which time steps were represented in each forecast ID directory, recording this information in a dictionary. Then, we implemented the logic of selecting the optimal file to open based on the time step we wish to represent. This was another point where the metadata provided was critical, since various dates in various date-time formats are reported for a given NetCDF file including weather forecast initialization, creation date of the NetCDF file, and smoke forecast initialization. The metadata headers are not explicitly explained on the FireSmoke portal. We extrapolated the meaning of the date entries by comparing them to similarly labeled dates available on the FireSmoke portal as shown in Figure \ref{fig:metaportal}. Implementation details for creating the sequence of hourly data are available in our supplemental materials.

We had to be intimately familiar with the metadata in NetCDF files to successfully select the optimal forecast for each time step. We leveraged metadata to dissect the time ranges represented across the collection of NetCDF files. Additionally, we had to confirm whether our assumption of selecting the latest forecast was appropriate or not. Given the various parameters needed to create a forecast and inherit uncertainty of forecast data, we were not sure if our assumption was correct. Communicating with UBC was the best way for us to confirm that our approach was appropriate.

\subsubsection{Debugging the Data Sequencing}

Implementing the logic of selecting the latest forecast required various attempts. We initially implemented a data picking procedure which immediately opened the NetCDF file for a given time step and copied the data to an array. However, for a search across 320 GB of data, this takes over a day to run.

Therefore, we switched to recording which NetCDF files opened successfully and the time steps they represent, which took only a few minutes to compute as we only read the metadata. Testing our sequencing procedure with these data structures rather than with the NetCDF files directly was more efficient.



\subsection{Managing Metadata}
Our final NetCDF file uses metadata from the original FireSmoke dataset and metadata that resulted from the data curation process. Table \ref{tab:dataset_attributes} shows the metadata that is included as variables in our NetCDF file.

\begin{table}[t]
    \centering
    \small  
    \setlength{\tabcolsep}{4pt}  
    \caption{Description of dataset attributes.}
    \label{tab:dataset_attributes}
    \begin{tabular}{p{2.5cm} p{5.5cm}}
        \toprule
        \textbf{Attribute} & \textbf{Description} \\
        \midrule
        PM25 & Concentration of PM2.5 for each time step, layer, row, and column in the spatial grid. \\
        TFLAG & Date and time of each data point. \\
        wrf\_arw\_init\_time & Time at which this prediction’s weather forecast was initiated. \\
        resampled & Indicates if the timestamp was resampled from a $381\times1041$ to $381\times1081$ grid. \\
        CDATE & Creation date of the data point (YYYYDDD format). \\
        CTIME & Creation time of the data point (HHMMSS format). \\
        WDATE & Date for which the weather forecast was initiated (YYYYDDD format). \\
        WTIME & Time for which the weather forecast was initiated (HHMMSS format). \\
        SDATE & Date for which the smoke forecast was initiated (YYYYDDD format). \\
        STIME & Time for which the smoke forecast was initiated (HHMMSS format). \\
        \bottomrule
    \end{tabular}
\end{table}

\subsubsection{Original Metadata}

We consulted external sources to understand the original metadata and considered changing the metadata format. For example changing date-time metadata from (YYYYDDD, HHMMSS) tuples to a more legible date-time string. However, we decided to preserve the metadata as it was originally represented. This preserves fidelity to the original dataset, ensuring consistency with our repurposed data, associated documentation, and the data creator’s best practices.

We instead provided demos and documentation showing how to read the metadata and transform it. This way, reusers can immediately know the meaning of the metadata and how to process it themselves.

\subsubsection{Derivative Metadata from Repurposing}

The uniformity of the NetCDF files enabled straightforward metadata extraction. Our final NetCDF file includes the original date and time information for each timestep and a variable indicating whether the data at a given timestep was resampled from a $341\times1041$ grid to a $341 \times 1081$ grid.



\subsection{Sharing Repurposed Data}
\begin{figure}[ht]
    \centering
    \includegraphics[width=\linewidth]{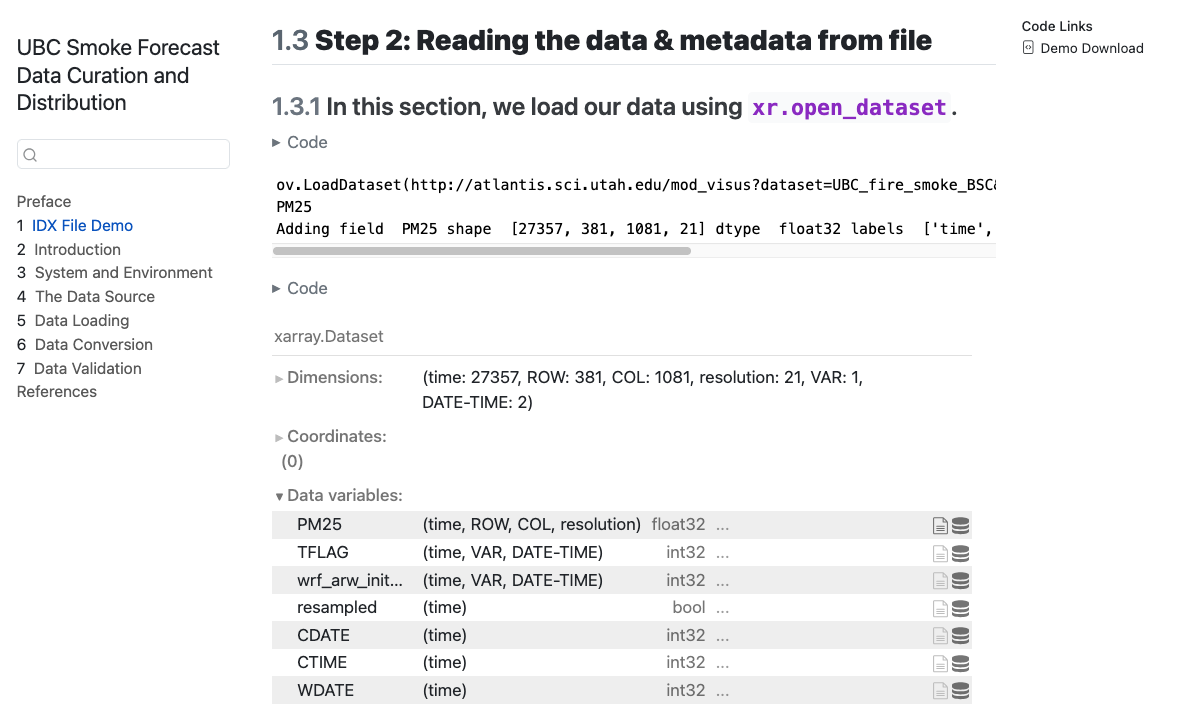}
    \caption{Jupyter Notebook demo for loading the NetCDF file of the repurposed dataset with Xarray, published with Quarto.}
    \label{fig:quarto1} 
\end{figure}

We published the documentation for the repurposed dataset in Quarto \parencite{quarto}. With Quarto we provide written documentation and Jupyter Notebook demos demonstrating the data repurposing process in this paper. Figure \ref{fig:quarto1} shows part of our demo. The documentation and NetCDF file containing the repurposed dataset are available in the supplemental materials. In the next section these materials are leveraged by our collaborators at the University of Calgary for their analysis on the impact of wildfire smoke on solar PV efficacy.

\section{Repurposed Data for Solar PV Efficacy}
\begin{figure}[ht]
    \centering
        \includegraphics[width=0.75\linewidth]{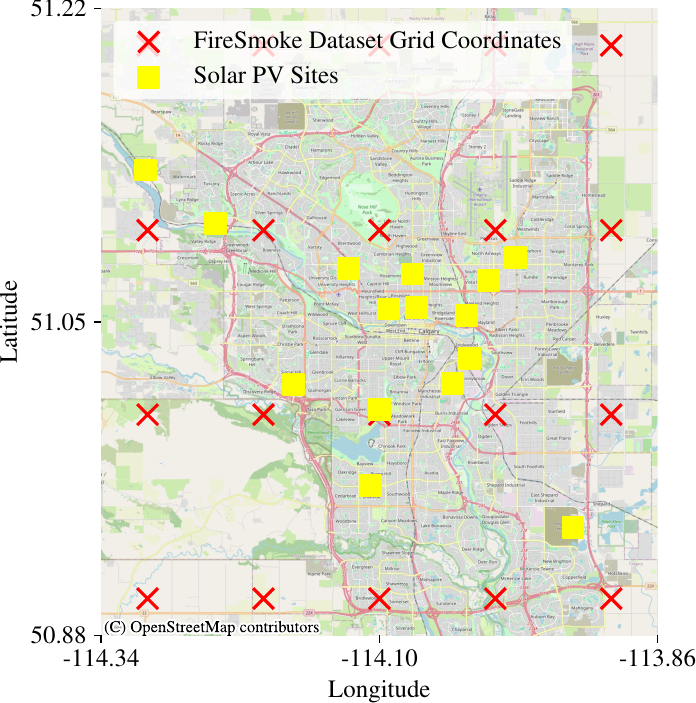}
        \caption{Map of Calgary with FireSmoke dataset grid coordinates and Solar PV Sites (coordinates in Web Mercator projection).}
        \label{fig:Calgary_map_no_p}
\end{figure}

\begin{figure}[ht]
    \centering
    \includegraphics[width=0.75\linewidth]{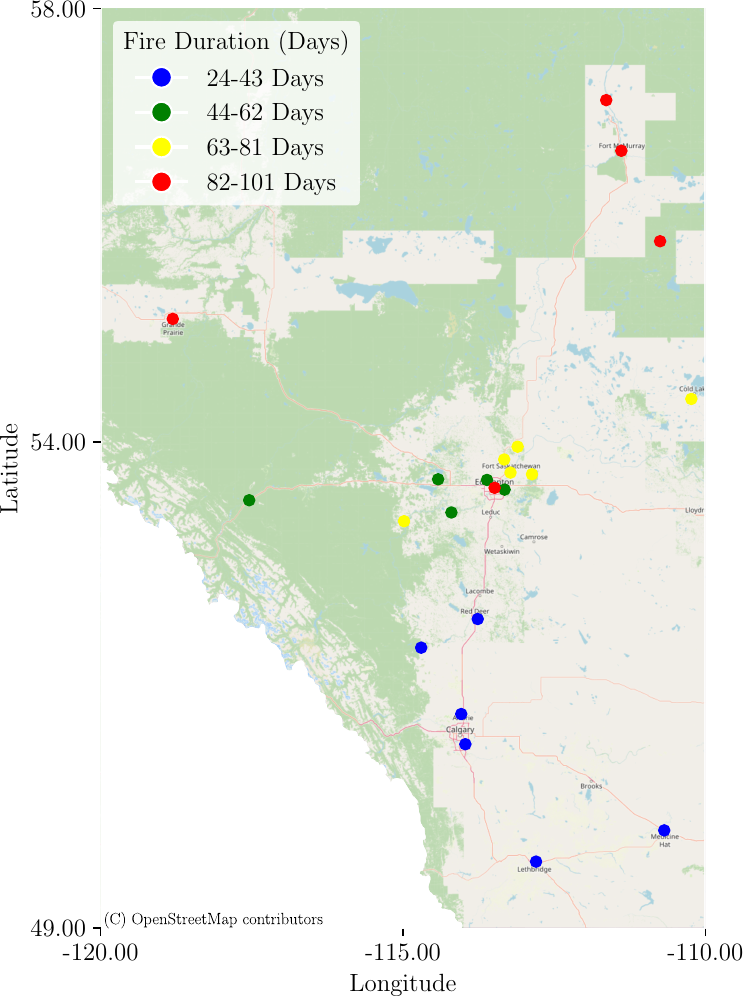}
    \caption{Number of days that select weather stations were impacted by wildfire smoke in Alberta in 2023 (coordinates in Web Mercator projection).}
    \label{fig:smoke_days} 
\end{figure}

Previous research encountered uncertainties by relying on distant ground-based PM2.5 monitoring stations, raising questions about the accuracy of assessing wildfire smoke impacts on solar PV performance. The repurposed dataset addresses this problem by utilizing a gridded PM2.5 data product that effectively ``fills the gaps" in regions lacking ground-based sensors. Data used in this Section belongs to the City of Calgary, located near the wildfire-prone Rocky Mountains in Alberta, Canada. In 2023, local weather stations reported that 24-43 days were impacted by wildfire smoke \parencite{AlbertaSmokyDays}. Figure \ref{fig:Calgary_map_no_p} shows solar PV sites and Figure \ref{fig:smoke_days} shows the number of smoky days observed at various locations across the province in 2023. Most solar PV sites are not close to a PM2.5 measurement station. However, with a geospatially comprehensive PM2.5 dataset, like the FireSmoke dataset, it is possible to have an estimate of a more localized value of PM2.5 for our analysis. We provide an analysis for one solar PV site at the University of Calgary.

\begin{figure}[ht]
    \centering
    \includegraphics[width=0.75\linewidth]{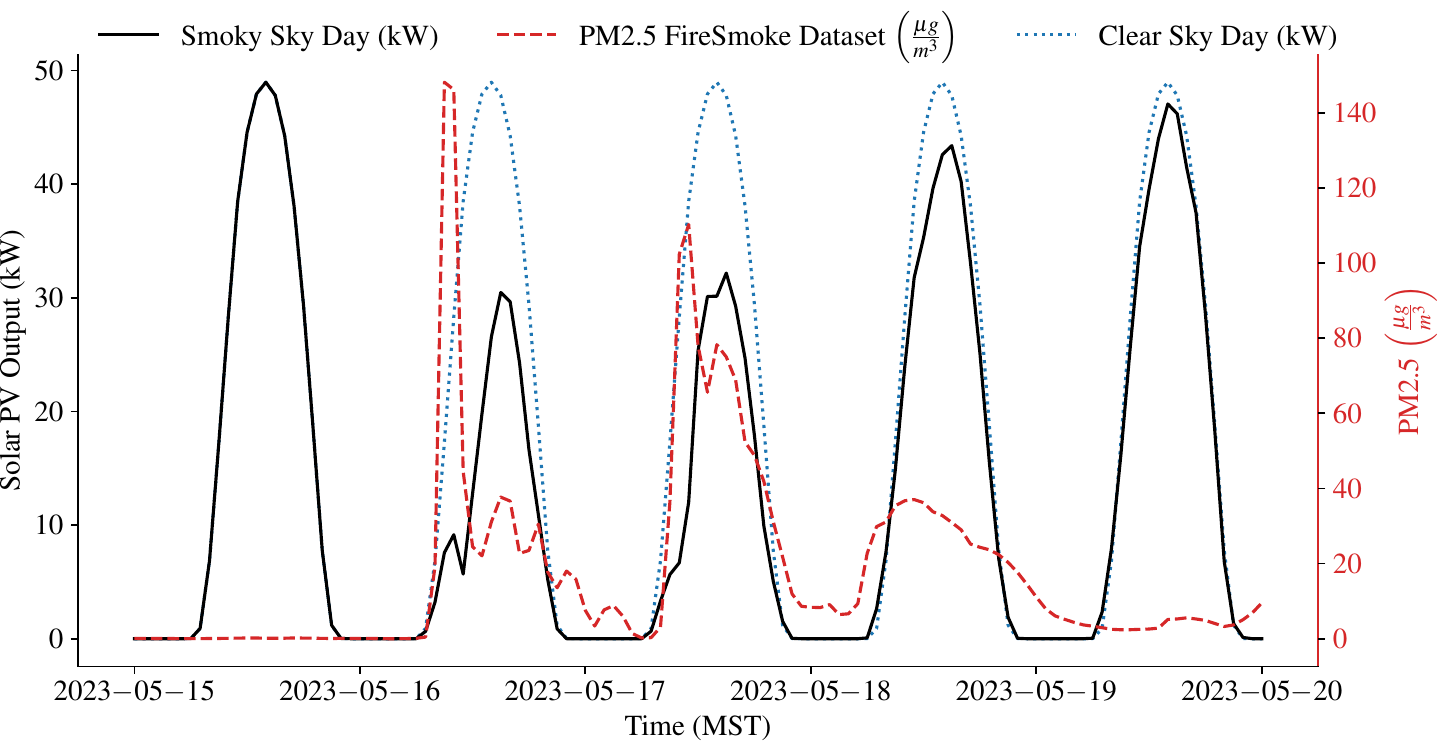}
    \caption{Solar Output and PM2.5 time series data.}
    \label{fig:smoke_PV_relationship} 
\end{figure}

\begin{figure}[ht]
    \centering
    \includegraphics[width=0.75\linewidth]{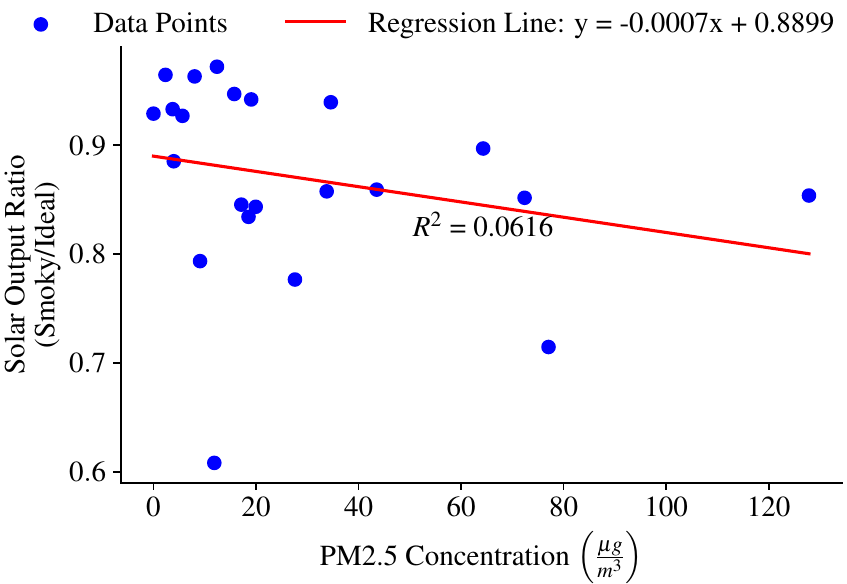}
    \caption{Solar Output Ratio vs. PM2.5 Concentration (for days with low cloud cover).}
    \label{fig:correlation_smoke_PV} 
\end{figure}



\begin{figure}[ht]
    \centering
    \begin{subfigure}[b]{0.48\linewidth}
        \centering
        \includegraphics[width=\linewidth]{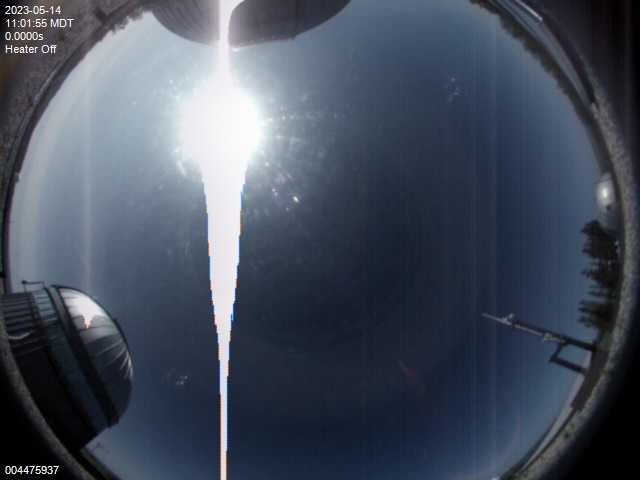}
        \caption{Skycam image of a clear sky on May 14, 2023.}
        \label{fig:Skycam image clear}
    \end{subfigure}
    \hfill
    \begin{subfigure}[b]{0.48\linewidth}
        \centering
        \includegraphics[width=\linewidth]{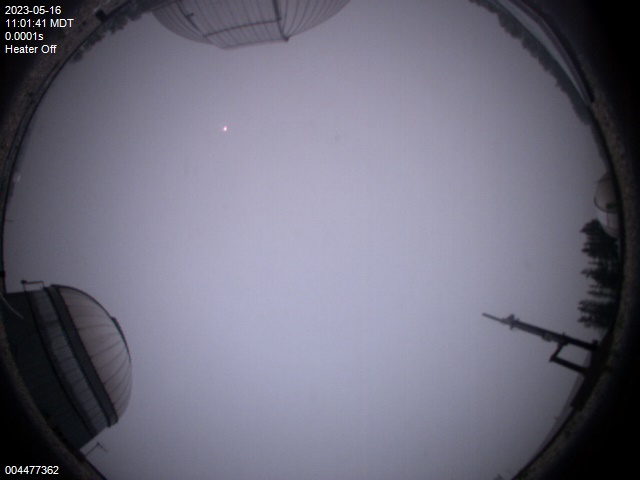}
        \caption{Skycam image of a smoky sky on May 16, 2023.}
        \label{fig:Skycam image smoky}
    \end{subfigure}
    \caption{Images taken with the RAO skycam.}
    \label{fig:Skycam comparison}
\end{figure}


\subsection{Datasets}
We collected solar PV production data from the 57 kW solar PV system installed at the University of Calgary main campus. The system became operational in 2018 and the energy production data was available at 15 minute intervals.

The repurposed FireSmoke dataset provides hourly PM2.5 concentration forecasts across 0.1° $\times$ 0.1° spatial grid cells. Figure \ref{fig:Calgary_map_no_p} shows the grid plotted over a map of Calgary and the solar PV sites distributed across the city.

Weather data was used for two metrics: cloud cover and presence of smoke. Cloud cover data was provided by \textcite{WeatherSource}. All the days where average cloud cover was higher than 20\% were eliminated to isolate the interference of clouds from analysis. Hourly observations on the presence of smoke were obtained from \textcite{ECCC_Historical_Data} (hereafter ECCC).

\subsection{Identifying Smoky and Clear-Sky Days}
Hourly ECCC weather reports were used to identify smoky days in Calgary, following the Manual of Surface Weather Observation Standards \parencite{ECCC_MANOBS_2023}. These observations were further corroborated using wildfire news reports and AllSky images from the Rothney Astrophysical Observatory (RAO), located south of Calgary. Figure \ref{fig:Skycam image clear} shows the sky on a clear day, while Figure \ref{fig:Skycam image smoky} shows a smoky day where the sun has been obscured by wildfire smoke.

We identified days in the solar PV dataset where daily production followed a smooth curve with respect to time. We classify these days as clear-sky days whereas days with a non-smooth curve are considered non-clear-sky days.

We used this data to identify dates where both smoky and clear-sky days occurred within a week of each other. In this paper, we provide an analysis for the week of May 15, 2023. Figure \ref{fig:smoke_PV_relationship} shows the solar PV energy output and PM2.5 concentrations from the repurposed FireSmoke dataset. We also plot expected solar PV output if the day had been a clear-sky day. Expected output is plotted to be the same output observed on an actual clear sky day earlier that week.

\subsection{FireSmoke Data Preparation}
We used the repurposed dataset to get hourly PM2.5 concentrations for the week of May 15, 2023. We used Xarray to slice the dataset for the dates of interest. However, choosing a PM2.5 value for a latitude and longitude not explicitly defined in the dataset was unclear. For instance, if we needed the PM2.5 value at a particular location, it was unclear which of the four surrounding grid points should be used. For this analysis, we selected the south-west grid point. We understood this point to represent the PM2.5 value for the corresponding square on the grid. However, upon sharing this approach with UBC, we were informed that interpolating the PM2.5 value based on the surrounding points is a better approach.

\subsection{Methodology}
We calculate the average solar PV output, PM2.5 concentration, and cloud cover percentage for each day, during peak production hours between 10:00 AM to 4:00 PM Mountain Daylight Time (MDT, UTC-6).

Linear regression was employed to quantify the relationship between daily average PM2.5 concentrations and solar output ratios. Figure \ref{fig:correlation_smoke_PV} illustrates the regression relationship. The ratio between the average solar PV output produced on a smoky day and the average energy produced on a similar clear-sky day was plotted on the y-axis. The ratio gives a better sense of comparison between energy production on smoky and clear-sky days, by reducing the impacts of other variables such as azimuthal angle.

\subsection{Results}
The regression results indicate reduction in solar energy output with an increase in PM2.5 concentration during days with minimal cloud cover. The regression analysis suggests a slightly negative correlation between PM2.5 concentration and solar output ratio, according to the equation $y = -0.0007x + 0.8899$.

These findings are generally consistent with previous research. Our results show a greater loss in solar PV output compared to \textcite{GILLETLY}. However, compared to \textcite{Korea}, the energy loss for the same concentration of PM2.5 is lower.

This preliminary study serves as a proof of concept, showcasing the viability of using repurposed data to analyze the relationship between wildfire smoke and solar PV output. Future research can focus on expanding the analysis to multiple solar PV sites across North America, to improve the generalizability and reliability of results. Furthermore, incorporating and evaluating sensor-based PM2.5 data from other sources near solar sites can provide a performance comparison of this repurposed dataset. Finally, we will use interpolation of the nearest grid points to find the PM2.5 value for an undefined latitude and longitude for future reanalyses.

\section{Lessons in Data Repurposing}

\subsection{Data Documentation}
The FireSmoke portal providing forecast data for view and download is a critical step in enabling access to the dataset for a broad audience. However, insufficient documentation led to tedious cross-checking when creating the historical dataset. Publishing the ranges of data available for download and the instructions for accessing older forecasts would easily solve this issue. The task of documenting data lineage and provenance is a cost and labor intensive task, especially for living datasets such as the FireSmoke dataset. Forecast parameters, publication schedules, and system maintenance frequently change and require attention from UBC. Without an incentive mechanism for data creators to create documentation for broad accessibility to their data, it is less likely to occur.

Data repurposing by more distant users is an opportunity to outsource tasks that do not frequently demand attention from data creators, such as identifying gaps in documentation. Direct collaboration enables the data creator and data reuser to identify these gaps and create documentation accordingly. Documentation by the data creator and data reuser will differ. However, documentation from both creators and reusers directly enable future reusers of a dataset to leverage this information.

\subsection{Data Sharing Infrastructure}
Sharing the resulting repurposed dataset on the cloud allows reusers to access the dataset anytime. Providing documentation and demos for using the repurposed dataset allowed us to transfer the knowledge needed for integrative reuse of the FireSmoke dataset. Additionally, a repurposed dataset and its associated documentation can be continually updated, allowing new knowledge or data processing to be immediately disseminated. As the research landscape continues to expand, the need for data sharing and knowledge infrastructures will increase as well.



\subsection{Data Understanding}
Atmospheric scientists and other domain sciences that frequently encounter geospatial data would likely know the best practices for interpreting and using the FireSmoke dataset. However, as the distance between the expertise and domain knowledge of the data creator and the data reuser grows, the risk of misusing the dataset also grows. 

In this case, examples of misuse include mishandling differing grid sizes in Section 4.2.3 and failing to interpolate using surrounding grid points in Section 5.3. The remedy for this was direct collaboration and guidance from UBC and accessing domain-level knowledge. Direct communication with the data creators provides the most actionable information. However, this is not scalable to a potentially large audience of data reusers for the FireSmoke dataset. 

Data creators have limited resources and must necessarily be selective about the level of assistance they can provide through direct collaboration and communication. Data creators will also not be around forever, meaning the time window for obtaining dataset-specific information needed to repurpose a dataset is short. Consequently, when data creators are not available, the next best knowledge source is the materials and resources that they provide. Knowledge exchange and gathering needed for repurposing would not be possible without support from the data creator and time and labor from the data reuser. Therefore, we find it critical to document knowledge exchange so it may persist for future users and scale to a larger audience.

We documented the repurposing process in our code and personal notes. We also frequently documented what assumptions we made and how we corrected them. We published the repurposing process on Quarto as a retrospective look at repurposing the data. We share the context and understanding that we believe future reusers would find useful in using our repurposed dataset or repurposing the FireSmoke dataset themselves. We also provide access to the associated files in implementing the repurposing, for reusers who may seek further details.

Although these knowledge infrastructures provide a medium for knowledge to be shared on a larger physical and temporal scale, there are limitations. Data repurposers have their own best practices and contexts. Therefore, the knowledge provided may not necessarily be enough to enable a new user to have the knowledge and context necessary to successfully repurpose a dataset. This is why, data repurposing is an iterative process not isolated to one instance, but across multiple. As reusers probe and transform the dataset and the knowledge provided about it, further knowledge exchange is sought. This leads to more knowledge that can be documented and disseminated.

\subsection{Cost Savings and Considerations}
Creating the FireSmoke dataset incurs significant computational and monetary costs for UBC. Their model is run on the Google Cloud Platform (GCP) using a minimal number of compute nodes to produce forecasts in a timely manner \parencite{chui2019producing}. Their model is run 4 times a day during the Canadian wildfire season. The yearly cost of this operation is about \$28,749.78 USD.

Extracting as much value as possible from the FireSmoke dataset is critical to enable broader audiences to benefit from the significant cost and effort incurred by UBC. We have shown that the significant cost and effort required to create the FireSmoke dataset does not directly translate to immediate data reusability. Data repurposing provides easier immediate access, both computationally and semantically to the FireSmoke dataset.

However, as with data creators, researchers need an incentive to put in time and effort to repurpose a dataset. The incentive for the authors at the University of Utah was to study the process of making the FireSmoke dataset more broadly available. For the authors at the University of Calgary, their incentive to use the repurposed dataset was to perform analyzes using data that is not available for their geographic regions of interest.

Repurposing a dataset for broader reuse is a secondary objective at best, and a liability at worst for domain science researchers. Unless scientific researchers are incentivized to provide public access and documentation, there is no driving force to enable data reusability beyond data sharing.

\section{Conclusion}\vspace{-0.25cm}
In this case study, we repurpose the FireSmoke dataset into a historical dataset for analyzing the effect of wildfire smoke on solar PV output. We highlight the challenges and opportunities in enabling data repurposing and reuse. By providing public access to the repurposed dataset and supporting documentation, we enable future integrative reuse of the FireSmoke dataset. This in turn will promote further knowledge exchange and help subsequent researchers overcome obstacles to data reusability.

\section{Supplemental Material}\vspace{-0.25cm}
All supplemental material for repurposing the FireSmoke dataset can be found at \url{https://github.com/sci-visus/NSDF-WIRED}. Supplemental material for the analysis of wildfire smoke on solar PV output can be found at \url{https://github.com/DGRILab/NSDF-WIRED}.

\section{Acknowledgements}
This paper is based on the work supported by the WIRED Global Center, jointly funded by the NSF GC grant 2330582 and NSERC grant ALLRP 585094-23. The work was also funded in part by NSF OAC award 2138811, NSF 
CI CoE Award 2127548, DOE SBIR Phase I DE-SC0026005, DOE SBIR Phase II DE-SC0017152, NSF POSE Phase I Award 2449026, NSF OISE award 2330582, the Advanced Research Projects Agency for Health (ARPA-H) grant no. D24AC00338-00, the Intel oneAPI Centers of Excellence at University of Utah, the NASA AMES cooperative agreements 80NSSC23M0013 and NASA JPL Subcontract No. 1685389.
Results presented in this paper were obtained in part using the Chameleon, Cloudlab, CloudBank, Fabric, and ACCESS testbeds supported by the National Science Foundation. This work was performed in part under the auspices of the DoE by LLNL under contract DE-AC52-07NA27344, (LDRD project SI-20-001).

\begingroup
\renewcommand*{\bibfont}{\small} 
\printbibliography
\endgroup

\end{document}